\documentclass[a4paper,floatfix,amsmath,amssymb,prb,twocolumn]{revtex4}
\usepackage{graphicx}
\usepackage{epsf}
\usepackage{dcolumn}
\usepackage{bm}
\usepackage{color}
\usepackage{caption}
\begin{document}
\newcommand{\newc}{\newcommand}
\newc{\mbf}{\mathbf}
\newc{\boma}{\boldmath}
\newc{\beq}{\begin{equation}}
\newc{\eeq}{\end{equation}}
\newc{\beqar}{\begin{eqnarray}}
\newc{\eeqar}{\end{eqnarray}}
\newc{\beqa}{\begin{eqnarray*}}
\newc{\eeqa}{\end{eqnarray*}}
\newc{\bd}{\begin{displaymath}}
\newc{\ed}{\end{displaymath}}


\title{Scaling of Rough Surfaces: Effects of Surface Diffusion on Growth and Roughness Exponents}
\author{Baisakhi Mal}
\affiliation{Department of Physics, Jadavpur University, Calcutta 
700 032, India}
\author{Subhankar Ray}
\email{sray@phys.jdvu.ac.in, sray_ju@rediffmail.com}
\affiliation{Department of Physics, Jadavpur University, Calcutta 
700 032, India}
\author{J. Shamanna}
\email{jlsphy@caluniv.ac.in}
\affiliation{Physics Department,University of Calcutta, 
Calcutta 700 009, India}

\begin{abstract}
Random deposition model with surface diffusion over several next
nearest neighbours is studied. The results agree with the results 
obtained by Family 
for the case of nearest neighbour diffusion [F. Family,
J. Phys. A 19(8), L441, 1986]. 
However for larger diffusion steps,
the growth exponent and the roughness exponent show 
interesting dependence on diffusion length.

\end{abstract}

\maketitle

\section{Introduction}
The growth of rough surfaces and interfaces plays a major role in numerous
phenomena of scientific interest and practical importance. 
There are many important
aspects to the growth of surfaces. Growing surfaces can evolve
into many forms. Flat, faceted, cusped and disorderly surfaces are familiar
forms. In addition, surfaces may also develop grooves, solid or hollow whiskers,
platelets, dendrites and other interesting structures. 
The problem of understanding
the physical processes that control these morphologies is a major
challenge with important practical implications. The
processes taking place in, on, and around disorderly interfaces are
also of interest for describing several microscopic phenomena.
A comprehensive study of growth surfaces 
is necessary to understand the internal structure of a wide range of objects.
The morphology, structure, other physical and chemical properties 
of growing interfaces
have been a subject of great interest 
in recent years. The study of these properties
finds applications in various physical, chemical and biological systems
and processes such as film growth by vapour deposition
\cite{thorn77,mess85,meak87,kar89,bales89}, 
bacterial growth\cite{fuji89,mat89,viscek90},
propagation of reaction fronts in catalyzed reactions \cite{albano97},
propagation of forest fires \cite{clar96} etc. 

Most of the simple models like Eden growth models \cite{eden},
ballistic deposition models \cite{vold}, solid-on-solid models
\cite{burton52}, reactive interface models \cite{kang93}, directed polymers
\cite{karzhang87}, polynuclear growth\cite{fran74, kash77, gil84}, 
directed percolation\cite{kin83} etc. were originally developed 
to simulate natural processes as diverse as growth of cell colonies,
sedimentation of colloids and crystallization of polymers. Some models
were also developed theoretically such as surface growth
with weak nonlinearity \cite{gate88} and surface growth with noise correlations
\cite{medina89}. Computer simulations
have played a major role in the development of better understanding
of surface growth phenomena under both equilibrium and non-equilibrium
conditions.

The simplest surface growth model is the random deposition model
where particles fall vertically on randomly chosen sites and 
are deposited on top of the respective columns \cite{barabasi}.
The roughness of the interface increases indefinitely without any
saturation. A modification of this model to better describe surface
growth phenomenon, namely, random deposition model with surface diffusion
was introduced by Family \cite{family86}. Here, the particles diffuse
or relax to the nearest neighbouring site. The roughness of the
interface increases initially and then saturates following
dynamic scaling relations.  

In this present work, we report the results of our study
of generalized random deposition models with 
surface diffusion for different diffusion lengths. In each of our models, 
the saturation widths and scaling exponents show a characteristic dependence
on the diffusion or relaxation length.

\section{Random deposition with surface diffusion}
A quantitative investigation of surface growth involves study
of interface width $W$, characterizing the roughness of the interface,
as a function of 
system size $L$ and the growth time $t$.
$W(L, t)$ is defined by the root-mean-square (rms) fluctuation
of the height of the interface. 
According to the dynamic scaling relation\cite{family85}, the
interface width $W(L, t)$ satisfies,
\beq\label{scale1}
W(L,t) = L^{\alpha}f(t/L^z)
\eeq
where $z = \alpha/\beta$ and 
$f$ is a scaling function satisfying $f(\infty) \sim \mbox{const}$ and
$f(x) \sim x^{\beta}$ for small $x$. Thus, for intermediate times 
$1 \ll t \ll \tau $, where $\tau$ is a model-dependent saturation time,
the interface width for a fixed system size $L$ has a power law
dependence on $t$,
\beq
W \sim t^{\beta}.
\eeq
And for $t\gg \tau$, it saturates to a time-independent value, 
$W_{sat}$, which scales with the system size $L$ as
\beq
W_{sat} \sim L^{\alpha}.
\eeq
The asymptotic scaling properties of the surface fluctuations 
of a given growth model are characterized completely by the 
growth exponent $\beta$ and roughness exponent $\alpha$ described above.

The model of random deposition with surface diffusion was 
introduced by Family\cite{family86}, in which a particle dropped 
in column $i$ sticks to the top
of the column $i$, $i+1$ or $i-1$ depending on which of the three columns
has the smallest height. In case, $i$, $i+1$ and $i-1$ columns have equal heights,
then the particle sticks to the top of either $i+1$ or $i-1$ column, with
equal probability. In the said study, it was 
claimed that there is no dependence of exponents on diffusion
length. In the same paper, the growth exponent $\beta$ was found to converge to 
a value of $0.25\pm 0.01$ asymptotically with increase in system size 
and the roughness exponent $\alpha$ was estimated to be $0.48\pm0.02$. 
It may be noted that the simulations were performed for relatively
small system sizes between $L=24$ and $L=384$.

In our present work, we study three variants of random deposition models with
surface diffusion. Each of these models reduce to Family's model
for single step diffusion ($N=1$). For larger system sizes 
between $L=128$ and $L=2048$, we observe significant and characteristic
dependence of various exponents on the allowed
number of diffusion steps $N$.

\subsection{Description of the models}
The models studied in this work can be described as follows.

\textbf{Model A:}
In this deposition model, particles are allowed to fall sequentially, 
one at a time, on randomly chosen sites on the growing surface
and are deposited on top of the selected columns. 
After this deposition, the particle remains at this site if it is the
only minimum among N-nearest neighbours on both sides.
Otherwise, the particle diffuses to the
nearest local minimum within the prescribed neighbourhood.
If the nearest minima on two sides are equidistant from the selected site,
the particle moves to either of these sites with equal probability.

However, in all of the models studied in this article,
the particle cannot jump over any local maximum to 
reach a local minimum within the prescribed neighbourhood. 

\begin{figure}[ht]
  \centering
  {\includegraphics[width=0.9\linewidth]{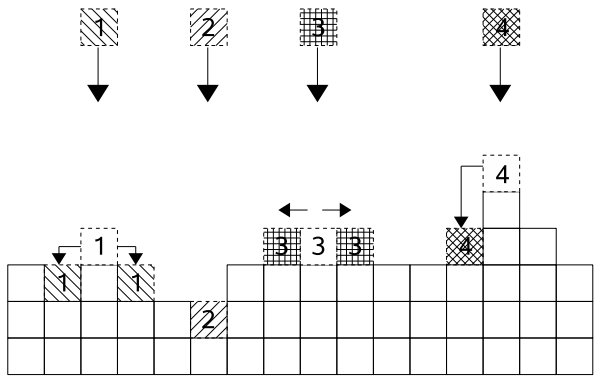}}
  \caption{Schematic diagram showing the rules of deposition for Model A. Probable sites for deposition are shown in different shades for different situations with maximum diffusion length $N=2$.}
  \label{famnear}
\end{figure}

\textbf{Model B:}
This model differs from Model A, in diffusion of the particle 
to the farthest local minimum within the prescribed neighbourhood.
If the farthest minima on two sides are equidistant from the selected site,
the particle moves to either of these sites with equal probability.
\begin{figure}[ht]
  \centering
  {\includegraphics[width=0.9\linewidth]{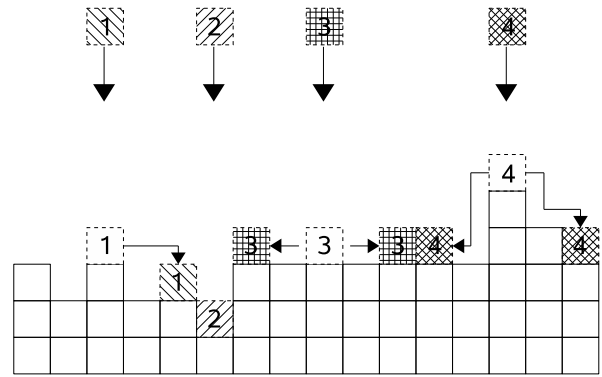}}
  \caption{Schematic diagram showing the rules of deposition for Model B. Probable sites for deposition are shown in different shades for different situations with maximum diffusion length $N=2$.}
  \label{famfar}
\end{figure}

\textbf{Model C:}
In this model, the particle moves to a minimum within N-nearest neighbours 
on both sides if this minimum is unique, otherwise, 
the particle moves randomly to
one of the several minima with equal probability.
\begin{figure}[ht]
  \centering
  {\includegraphics[width=0.9\linewidth]{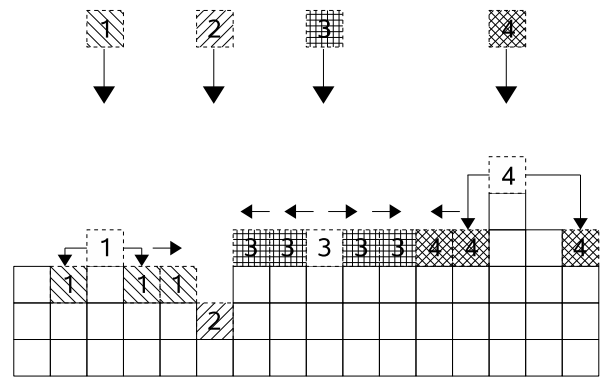}}
  \caption{Schematic diagram showing the rules of deposition for Model C. Probable sites for deposition are shown in different shades for different situations with maximum diffusion length $N=2$.}
  \label{difeq}
\end{figure}

In all the above variants of random deposition model with surface diffusion, 
two random number generators were used, one for selecting the site on the
growing surface and, another for depositing the particle with equal
probability when there are multiple minima. These two random number generators
are completely independent and 
uncorrelated to each other. These random number generators are based on
linear congruential method \cite{knuth,sedge}. 

\section{Results}
\begin{figure}[ht]
  \centering
  {\includegraphics[width=0.9\linewidth]{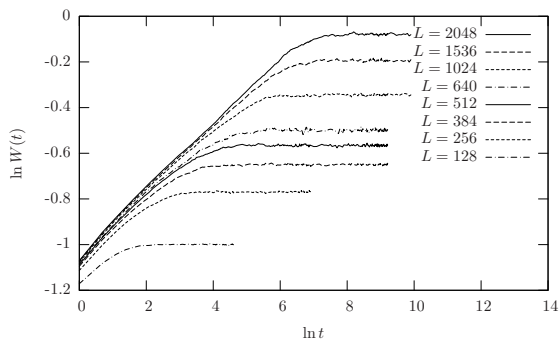}}
  \caption{$\ln W(t)$ vs $\ln t$ showing the growth and
      saturation regions for Model A, $L=128$ to $L=2048$ and $N=15$.}
  \label{nearall}
\end{figure}

\begin{figure}
  \centering
  {\includegraphics[width=0.9\linewidth]{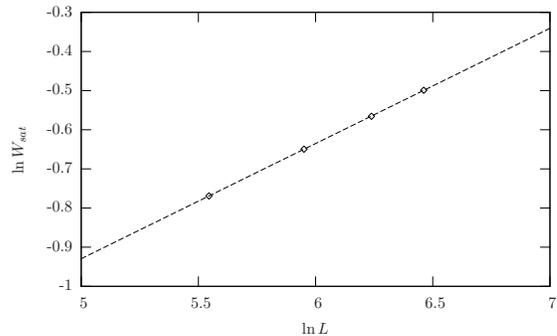}}
  \caption{$\ln W_{sat}$ vs $\ln L$ with straight line fit for $\alpha$.}
  \label{alphafit}
\end{figure}

The interface widths $W(L,t)$ were computed for different time-steps
and for different system sizes between $L=128 $ and $L=2048$.
When presented graphically in log-log scale, these graphs show
distinct growth and saturation regions. One such graph for Model A,
for system sizes between $L=128 $ and $L=2048$ with diffusion step
$N=15$ is shown in Fig.\ref{nearall}. 
The saturation widths $W_{sat}$ for various
system sizes and diffusion lengths were calculated for all the models
by means of a straight line fit of $\ln W_{sat}$ versus $\ln L$. One such
characteristic linear fit for finding the roughness exponent $\alpha$ 
is shown for Model A with diffusion step $N=15$ in Fig.\ref{alphafit}. 
The related data for the same is shown in Table \ref{alptable}.

\begin{table}[ht]
  \caption{$W_{sat}$ as function of system size $L$ for Model A
    with diffusion step $N = 15$.}
  \label{alptable}
  \begin{tabular}{|@{\;\;}c@{\;\;}|@{\;\;}c@{\;\;}|@{\;\;}c@{\;\;}|}
    \hline
    L & $\ln L$ & $\ln W_{sat}$ \\
    \hline
    256 & 5.5452 & -0.7690 \\
    384 & 5.9506 & -0.6496 \\
    512 & 6.2383 & -0.5656 \\
    640 & 6.4615 & -0.4991 \\
    \hline
  \end{tabular}
\end{table}%

\begin{figure}
  \centering 
  {\includegraphics[width=0.9\linewidth]{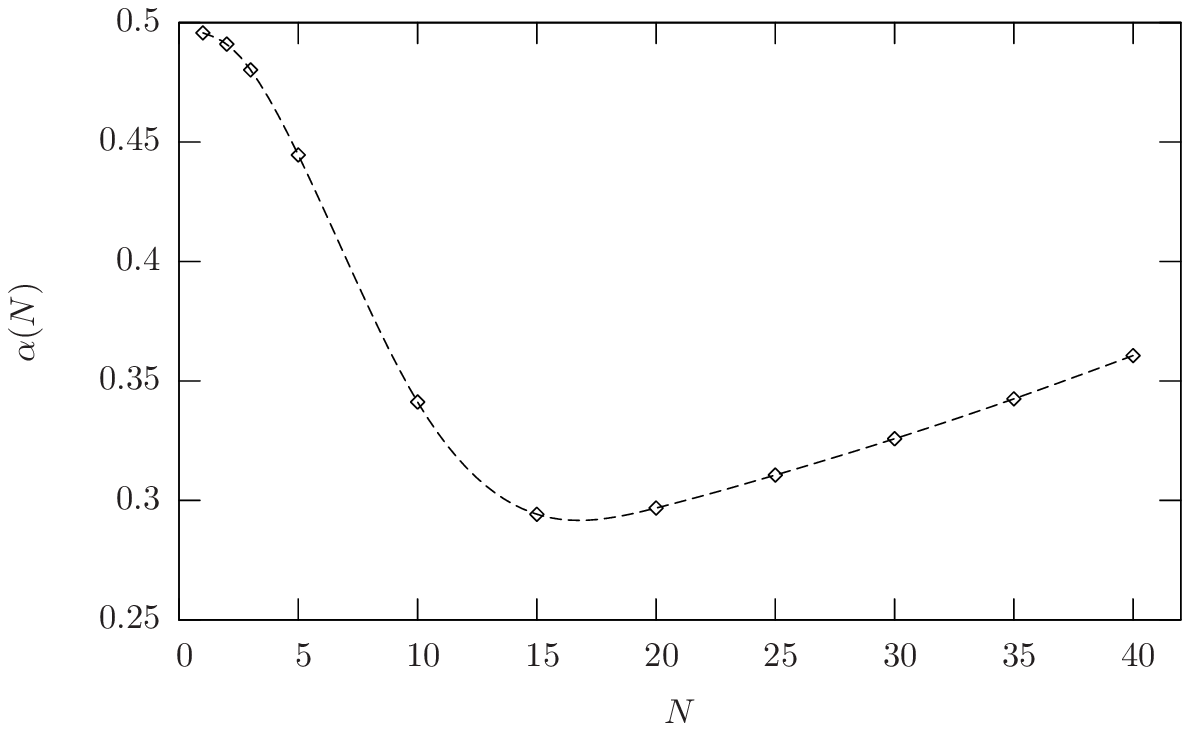}}
  \caption{Dependence of roughness exponent $\alpha$ on number of
    diffusion steps $N$ for Model A. }
  \label{nearal}
\end{figure}

\begin{figure}
  \centering
  {\includegraphics[width=0.9\linewidth]{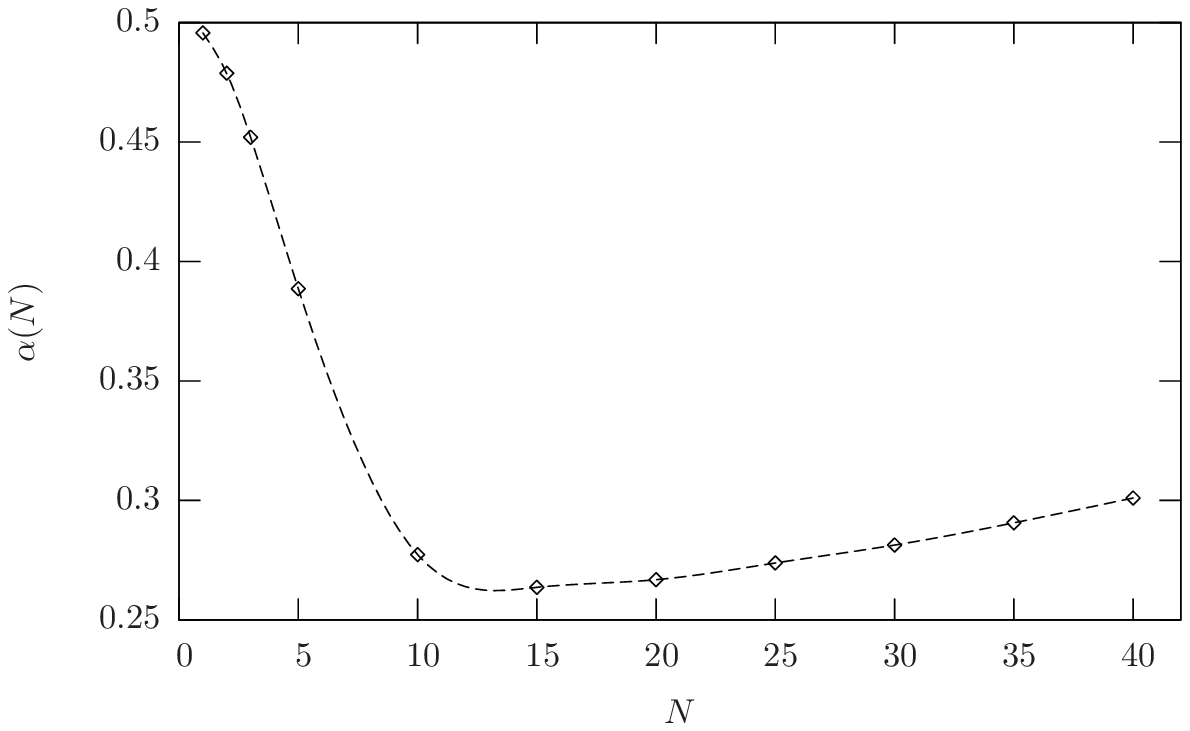}}
  \caption{Dependence of roughness exponent $\alpha$ on number of
    diffusion steps $N$ for Model B. }
  \label{faral}
\end{figure}

\begin{figure}
  \centering 
  {\includegraphics[width=0.9\linewidth]{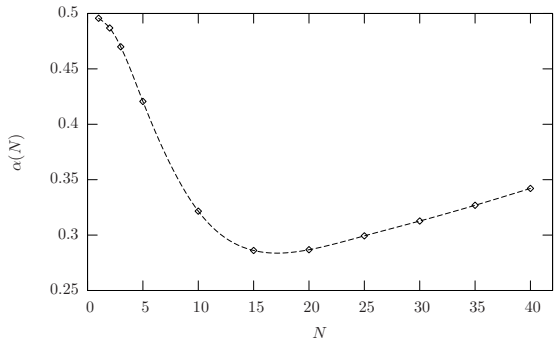}}
  \caption{Dependence of roughness exponent $\alpha$ on number of
    diffusion steps $N$ for Model C.}
  \label{eqal}
\end{figure}

The roughness exponent $\alpha$ shows an interesting 
dependence on the number of diffusion
steps $N$ in all of the models studied. It shows
a sharp decrease initially reaching a minimum around $N=15$.
With further increase in diffusion steps,
it increases slowly. The relevant data is given in Table \ref{alphatab}
and the dependence is shown in Fig.\ref{nearal}, \ref{faral} and \ref{eqal}. 
\begin{table}
  \centering
  \caption{Roughness exponent $\alpha$ 
    for various diffusion steps.}
  \label{alphatab}
  \begin{tabular}{|@{\;\;}c@{\;\;}|@{\;\;}c@{\;\;}|@{\;\;}c@{\;\;}|@{\;\;}c@{\;\;}|}
    \hline
    N & Model A & Model B & Model C \\
    \hline
    1 &  0.4957 & 0.4957 & 0.4957\\
    2 &  0.4910 & 0.4788 & 0.4869\\
    3 &  0.4802 & 0.4520 & 0.4699\\
    5 &  0.4446 & 0.3886 & 0.4206\\
    10 &  0.3412 & 0.2733 & 0.3216\\
    15 &  0.2942 & 0.2636 & 0.2861\\
    20 &  0.2968 & 0.2668 & 0.2868\\
    25 &  0.3106 & 0.2738 & 0.2993\\
    30 &  0.3258 & 0.2813 & 0.3127\\
    35 &  0.3425 & 0.2906 & 0.3269\\
    40 &  0.3606 & 0.3010 & 0.3421\\
    \hline
  \end{tabular}
\end{table}

\begin{figure}
  \centering 
  {\includegraphics[width=0.9\linewidth]{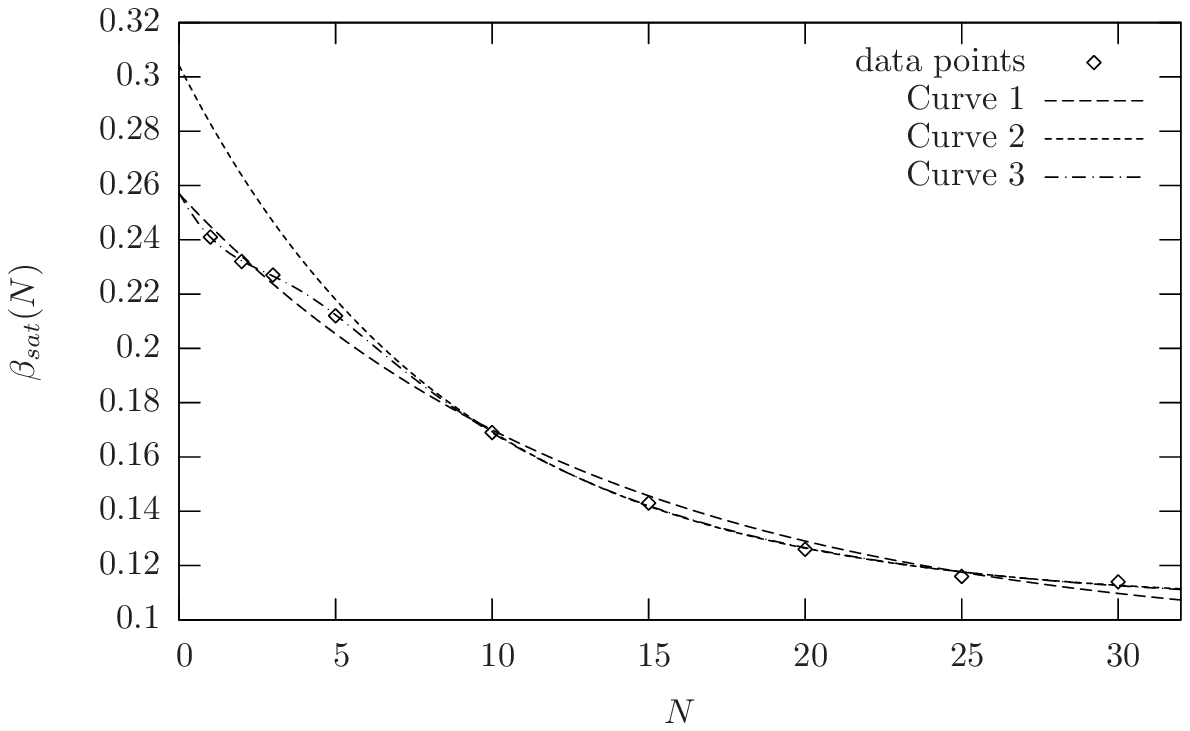}}
  \caption{Dependence of growth exponent $\beta$ on number of diffusion
    steps $N$ for Model A.}
  \label{nearbeta}
\end{figure}

\begin{figure}
  \centering 
  {\includegraphics[width=0.9\linewidth]{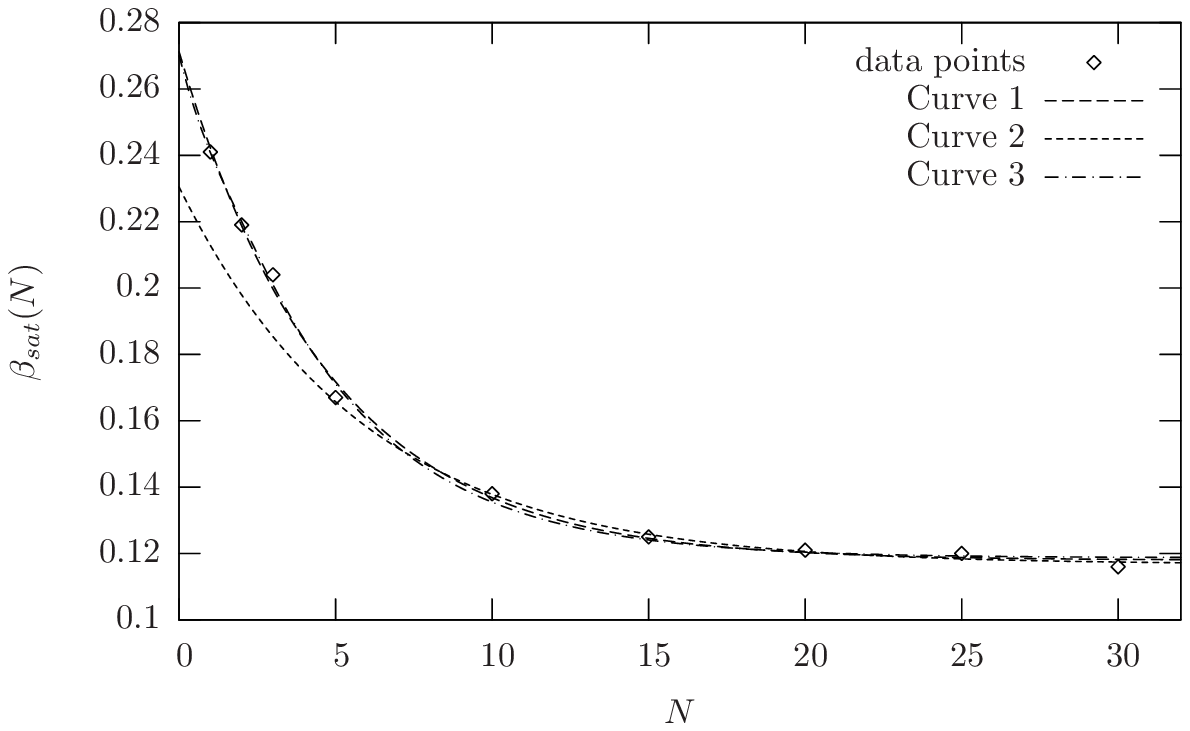}}
  \caption{Dependence of growth exponent $\beta$ on number of diffusion
    steps $N$ for Model B.}
  \label{farbeta}
\end{figure}

\begin{figure}
  \centering 
  {\includegraphics[width=0.9\linewidth]{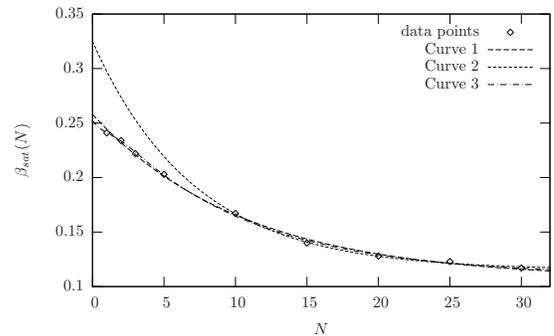}}
  \caption{Dependence of growth exponent $\beta$ on number of diffusion
    steps $N$ for Model C.}
  \label{eqbeta}
\end{figure}

The growth exponent $\beta$ was calculated from the slope of 
the growth region in $\ln W(t)$ versus $\ln t$ graph. 
With increase in system size, the growth exponent approaches 
a saturation value $\beta_{sat}$. It is further observed that
this saturation value also shows a strong 
dependence on the number of diffusion steps $N$.
With increase in $N$, $\beta_{sat}$ asymptotically approaches a steady 
value $\beta_{steady}$.
The relevant data showing this behaviour is given in Table \ref{betatab}
and the dependence is shown in Fig.\ref{nearbeta}, \ref{farbeta} and \ref{eqbeta}. 

To study the saturation behaviour $\beta_{sat}$ with $N$, 
two empirical forms of dependence were
assumed and data were fitted to these non-linear forms.
\beq\label{fit1}
\beta_{sat} = b_0 + b_1 e^{-\nu N}
\eeq
\beq\label{fit2}
\beta_{sat} = b_0 + b_1 e^{-\nu N} - b_2 e^{-\mu N^2}
\eeq
where $\beta_{steady} = b_0$ for the corresponding expression.
The values estimated for $b$'s are shown in Table \ref{curve}.
Both the above forms show saturation with increase in 
diffusion length. As the fit with single exponential function,
Eq.(\ref{fit1}) was not very faithful over the entire range of values
of diffusion length, particularly for Model A, a fit was also made 
to include data only corresponding to higher values of diffusion length.
The empirical form of Eq.(\ref{fit2}) however,
shows an improved fit for Model A over the entire range of 
data points.

\begin{table}
  \centering
  \caption{Saturation growth exponent $\beta_{sat}$ for 
various diffusion steps.}
  \label{betatab}
  \begin{tabular}{|@{\;\;}c@{\;\;}|@{\;\;}c@{\;\;}|@{\;\;}c@{\;\;}|@{\;\;}c@{\;\;}|}
    \hline
    N & Model A & Model B & Model C \\
    \hline
    1 & 0.241 & 0.241 & 0.241 \\
    2 & 0.232 & 0.219 & 0.234 \\
    3 & 0.227 & 0.204 & 0.222 \\
    5 & 0.212 & 0.167 & 0.203 \\
    10 & 0.169 & 0.132 & 0.167 \\
    15 & 0.143 & 0.125 & 0.140 \\
    20 & 0.126 & 0.121 & 0.128 \\
    25 & 0.116 & 0.119 & 0.123 \\
    30 & 0.114 & 0.116 & 0.117 \\
    \hline
  \end{tabular}
\end{table}

\begin{table}
    \caption{Values of coefficients $b$ 
for different curves fitted to $\beta_{sat}$ versus $N$ data}
    \label{curve}
  \centering
  \begin{tabular}{|c|c|c|c|c|}
      \hline
      Fitted curves for &  & Model A & Model B & Model C \\
      different models &&&& \\
      \hline
      All data fitted to Eq.(\ref{fit1}) & $b_0$ &0.0925  &0.1179 &0.1069 \\
      (Curve 1) & $b_1$ &0.1644 &0.1532 &0.1511 \\
      \hline
      Large diffusion length & $b_0$ &0.1063 &0.1167 &0.1151 \\
      data fitted to Eq.(\ref{fit1}) & $b_1$ & 0.1979 &0.1138 &0.2099 \\
      (Curve 2) &&&& \\
      \hline
      All data fitted to Eq.(\ref{fit2}) & $b_0$ &0.1056 &0.1187 &0.1100 \\
      (Curve 3) & $b_1$ &0.1912 &0.1636 &0.1555 \\
      & $b_2$ &0.0396 &0.0123 &0.0135 \\
      \hline
  \end{tabular}
\end{table}

\section{Conclusion}
It was claimed in the study by Family \cite{family86} that the saturation 
widths and the exponents were independent of the diffusion length.
However, in our study, we have found that both the growth exponent $\beta$
and the roughness exponent $\alpha$ depend on the diffusion length. 
It may be noted that in the said reference, the 
simulations were carried out for relatively small
system sizes between $L=24$ and $L=384$. The study was therefore limited to
smaller diffusion lengths, and perhaps the dependence on number of 
diffusion steps $N$ was missed.

When the system size is large,
the extent of diffusion of a deposited particle is limited 
only by the diffusion
length. As the diffusion length is increased, faster and more uniform
smoothening of the interface is observed. As a result, $W_{sat}$ decreases
with both system size $L$ and time $t$,
thereby decreasing the exponents $\alpha$ and $\beta$. As a 
mechanism we propose that, as the
diffusion length is further increased, the most jagged boundaries have
smoothened to give a series of wide shallow valleys separated by crests.
When the diffusion length becomes comparable to the width of these shallow
valleys, a depositing particle has a higher probability
of encountering the boundary of these valleys, i.e., the crests, before 
reaching the diffusion limit. 
As none of the models studied here allow surmounting
any hill or crest while approaching a minimum on the other side, the
diffusion of a particle gets restricted if it encounters a valley boundary
before reaching its diffusion limit. This happens predominantly when the
valley width is comparable with the diffusion length. In these cases, 
the dynamics is dominated by the existence of valley boundaries rather 
than the diffusion length. This is observed as a saturation in the 
growth exponent $\beta$ and decrease in roughness exponent 
$\alpha$ followed by a slow increase. This behaviour needs to be
investigated in further detail.


\begin{thebibliography}{99}
\bibitem{thorn77} J. A. Thornton, Ann. Rev. Mater. Sci. {\bf 7}, 239 (1977).
\bibitem{mess85} R. Messier and J. E. Yehoda, J. Appl. Phys. {\bf 58}, 3739 (1985).
\bibitem{meak87} P. Meakin, J. Phys. A {\bf 20}(16), L1113 (1987).
\bibitem{kar89} R. P. U. Karunasiri, R. Bruinsma and J. Rudnick, Phys. Rev. Lett. {\bf 62}(7), 788 (1989); Erratum: Phys. Rev. Lett. {\bf 62}(23), 2767 (1989).
\bibitem{bales89} G. S. Bales and A. Zangwill, Phy. Rev. Lett. {\bf 63}(6), 692 (1989).
\bibitem{fuji89} H. Fujikawa and M. J. Matsushita, J. Phys. Soc. Japan {\bf 58}, 3875 (1989).
\bibitem{mat89} T. Matsuyama, M. Sogawa and Y. Nakagawa, FEMS Microbiology Lett. {\bf 61}(3), 243 (1989)
\bibitem{viscek90} T. Viscek, M. Cserzo and V. K. Horvath, Physica A {\bf 167}(2), 315 (1990).
\bibitem{albano97} E. V. Albano, Phy. Rev. E {\bf 55}, 7144 (1997).
\bibitem{clar96} S. Clar, B. Drossel and F. Schwabl, J. Phys.: Condens. Matter {\bf 8}(37), 6803, (1996).
\bibitem{eden} M. Eden, {\it Symposium on Information Theory in Biology}, Pergamon press, New York (1958); M. Eden, {\it Proc. 4th Berkeley Symposium on Mathematics Statistics and Probability}, Vol 4 (1961).
\bibitem{vold} M. J. Vold, J. Colloid Sci. {\bf 14}(2),168 (1959); J. Phys. Chem. {\bf 63}(10), 1608 (1959); J. Phys. Chem. {\bf 64}(11), 1616 (1960).
\bibitem{burton52} W. K. Burton, N. Cabrere and F. C. Frank, Phil. Trans. Roy. A Soc {\bf 243}(10) , 299 (1951).
\bibitem{kang93} H. C. Kang and W. H. Weinberg, Phys. Rev. E {\bf 47}(3), 1604 (1993).
\bibitem{karzhang87} M. Kardar and Y. C. Zhang, Phy. Rev. Lett. {\bf 58}(20), 2087 (1987).
\bibitem{fran74} F. C. Frank, J. Crystal Growth {\bf 22}(3), 233 (1974).
\bibitem{kash77} D. Kashchiev, J. Crystal Growth {\bf 40}(1), 29 (1977).
\bibitem{gil84} G. H. Gilmer, J. Crystal Growth {\bf 49}(3), 465 (1984).
\bibitem{kin83} W. Kinzel, {\it Percolation Structures and Processes}, eds. G. Deutcher, R. Zellen and J. Adler, (Hilger, Bristol),(1983).
\bibitem{gate88} D. J. Gates and M. Westcott, Proc. Roy. Soc. {\bf 416}(1851), 443 (1988); Proc. Roy. Soc. {\bf 416}(1851), 463 (1988).
\bibitem{medina89} E. Medina, T. Hwa, M. Kardar and Y. C. Zhang, Phys. Rev. A {\bf 39}(6), 3053 (1989).
\bibitem{barabasi} A. L. Barabasi and H. E. Stanley, {\it Fractal Concepts in Surface Growth}, Cambridge University Press (1995).
\bibitem{family86} F. Family, J. Phys. A {\bf 19}(8), L441 (1986).
\bibitem{family85} F. Family and T. Viscek, J. Phys. A {\bf 18}(2), L75 (1985).
\bibitem{knuth} D. E. Knuth, {\it The Art of Computer Programming, Volume 2, Seminumerical Algorithms}, Addison-Wesley, Reading Mass. (1981).
\bibitem{sedge} R. Sedgewick, {\it Algorithms in C++}, Addison-Wesley, Reading Mass. (1992). 
\end{thebibliography}
\end{document}